\documentclass[aps,prl,twocolumn,showpacs,amsmath,amssymb]{revtex4}

\usepackage{graphicx}
\usepackage{dcolumn}
\usepackage{bm}

\begin{document}

\def\bs{\boldsymbol}

\preprint{Faraday}

\title{Broken symmetries and pattern formation in two-frequency forced Faraday waves}

\author{Jeff Porter}
\email{jport@northwestern.edu}
\author{Mary Silber}
\affiliation{Department of Engineering Sciences and Applied Mathematics,
Northwestern University, Evanston, Illinois 60208}

\date{\today}

\begin{abstract}
We exploit the presence of approximate (broken) symmetries to obtain general scaling laws 
governing the process of pattern formation in weakly damped Faraday waves.   Specifically, we 
consider a two-frequency forcing function and trace the effects of time translation, time reversal 
and Hamiltonian structure for three illustrative examples: hexagons, two-mode superlattices, and 
two-mode rhomboids.  By means of explicit parameter symmetries, we show how the size of various 
three-wave resonant interactions depends on the frequency ratio $m$:$n$ and on the relative 
temporal phase of the two driving terms.  These symmetry-based predictions are verified for 
numerically calculated coefficients, and help explain the  results of recent experiments.
\end{abstract}

\pacs{47.54.+r, 05.45.-a, 47.35.+i, 47.20.Gv}

\maketitle

Symmetry arguments play a central role in our understanding of pattern formation in 
nonlinear, nonequilibrium systems. They explain, for example, the ubiquity of simple 
patterns such as hexagons, squares and stripes observed in widely disparate systems, both in 
nature and in the laboratory \cite{CH93}. Group theory provides the natural language for 
describing and classifying patterns and, when combined with bifurcation theory, it determines 
which patterns may result from a symmetry-breaking instability \cite{GS85}. It  is by now 
standard practice to invoke symmetry arguments to determine which nonlinear terms are present 
in the universal amplitude equations that govern the pattern formation processes near onset. The 
detailed physics is manifest in these amplitude equations only through the numerical values of the 
coefficients of the nonlinear terms, yet the computation of these coefficients is often an arduous 
task that must be carried out numerically.

In this letter we develop symmetry arguments in a somewhat nonstandard way, 
using {\it parameter symmetries} (symmetries which act on both amplitudes and 
parameters; see e.g.~\cite{S88}) to determine the dependence of the most important 
{\it coefficients} in the amplitude equations on the parameters of the forcing function and 
on the damping.  The specific problem we consider is that of gravity-capillary wave 
excitation on the free surface of a fluid subjected to the (vertical) forcing function
\begin{align}
F(t) &= \frac{1}{2}(f_m e^{im\omega t} + f_n e^{i n \omega t} + cc.) \\ 
& \equiv |f_m|\cos(m\omega t + \phi_m) + |f_n|\cos(n\omega t + \phi_n),\nonumber
\end{align}
where $m$ and $n$ are relatively prime.  This system, with its wealth of readily tuneable 
control parameters, has been a rich source of intriguing patterns \cite{EF94,KPG98,AF98,AF00,AF01}.   
Although a variety of nonlinear interactions influence the pattern selection process, resonant triads 
are especially important because they arise at second order in a weakly-nonlinear expansion and 
lead to a strong phase coupling between the three waves involved - they have been implicated in many 
Faraday wave pattern studies, both experimental \cite{AF98,AF00,AF01} and theoretical 
\cite{ZV97,STS00}.   This letter uses general symmetry arguments applied to several such three-wave 
interactions to shed some light on the role of the control parameters ($|f_m|,\ |f_n|,
\ m\omega,\ n\omega,\ \phi_m,\ \phi_n$) in the pattern formation process.

Symmetry considerations, both spatial and temporal,  have already gone a long way toward 
elucidating the general mechanisms by which different patterns arise.  For example, when a 
{\it single} frequency $\omega$ is used, a sufficiently large forcing excites standing 
waves (SW) with primary frequency $\omega/2$.  After one period $T$ of the driving, the SW 
eigenmodes $\psi_j$ are exactly out of phase and, due to the symmetry $t \rightarrow t+T,\ \psi_j 
\rightarrow -\psi_j$, are barred from quadratic (i.e.~three-wave) interactions. The addition of a 
second commensurate frequency (when not a multiple of the first) breaks this particular discrete 
translation symmetry and replaces it with one on a {\it longer} timescale: $t\rightarrow t+p\,T$.   
If $p$ is even this weaker symmetry constraint permits three-wave interactions by recasting the 
$\psi_j$ as {\it harmonic} modes.   Despite the success of this argument in explaining the emergence  
of hexagons when a second frequency is added \cite{EF94} it fails to predict either the size of the 
symmetry-breaking effect or its dependence on the control parameters.  

In this letter we take advantage of nearby (approximate) symmetries to make further progress.  
The main idea is that, for weakly-damped small-amplitude dynamics, the real physical system  
(damped forced SW) is intimately related to an underlying system (undamped unforced traveling 
waves (TW)) that is subject to more stringent symmetry requirements: invariance under 
{\it arbitrary} time translations $t \rightarrow t+\tau$, and time reversal 
$t \rightarrow -t$.  If the undamped but forced problem is Hamiltonian the system is still further 
constrained.  Although the addition of forcing and damping destroys these symmetries, 
their influence persists.  In fact, the broken symmetries can be replaced by corresponding 
parameter symmetries which remain intact for {\it any} damping and forcing.  It is only to 
obtain useful scaling laws that we must restrict ourselves to {\it small} damping and forcing.

\begin{figure}[h]
\centerline{\includegraphics[width=2.7in]{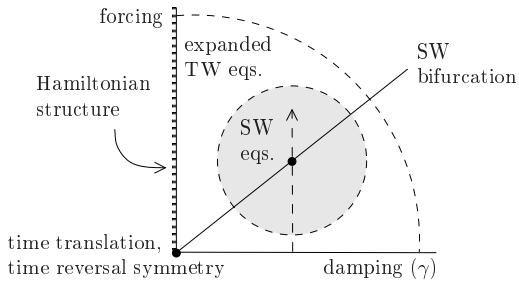}}
\caption{Relationship of SW equations to underlying symmetries.  
The origin denotes systems with time translation and time reversal symmetry.  
The undamped problem is Hamiltonian.  SW bifurcate from the flat state along 
the solid line, while the dashed vertical line illustrates a realistic experimental path.}
\label{fig:unfold}
\end{figure}

The overall strategy is illustrated in Fig.~\ref{fig:unfold}.  
First, we express physical variables such as surface height $h({\bs x},t)$ in terms of 
the TW eigenfunctions of the undamped unforced problem:
\begin{align}
h({\bs x},t) = \sum_{j=1}^N\sum_{\pm} Z_j^\pm (t)\,e^{i({\bs k_j}\cdot 
{\bs x}\pm \Omega_j t)} + cc. + ...\,.
\label{eq:h}
\end{align} 
In all cases $N$ is finite (we consider only periodic patterns). Evolution equations for the TW
amplitudes $Z_j^{\pm}$ must respect the appropriate spatial symmetries, which depend on 
the planform considered, and the following parameter symmetries:
\begin{align}
&T_\tau \!: Z_j^{\pm} \rightarrow e^{\pm i \Omega_j \tau}Z_j^{\pm}\!,\:
(f_m,f_n) \rightarrow ( f_m e^{im\tau}\!, f_n e^{in\tau}), \nonumber \\ 
&\kappa : Z_j^{\pm}\! \leftrightarrow Z_j^{\mp},\: (t,\gamma) \rightarrow\! -(t,\gamma),\: 
(f_m,f_n) \rightarrow (\bar{f}_m,\bar{f}_n).
\label{eq:tempsyms}
\end{align}
Here $\gamma = 2 \nu k^2/\omega$ is a dimensionless damping parameter ($\nu$ denotes 
viscosity and $k(\omega)$ is a characteristic wavenumber).  Next, we assume the TW equations 
can be expanded in powers of $\gamma$, $f_m$ and $f_n$ and so determine their  form for small 
damping and forcing.  Lastly, these expanded TW equations are reduced at the primary 
bifurcation through a center manifold reduction to yield the equations describing SW near onset.  
The procedure outlined above can be applied to numerous patterns 
and we shall look briefly at three: hexagons, two-mode superlattices \cite{AF00}, and two-mode 
rhomboids \cite{AF98}.  We test the predicted dependence of SW coefficients by 
numerically calculating them from the Zhang-Vi\~nals model \cite{ZV97}, a quasipotential 
formulation of the Faraday problem appropriate for a deep layer of nearly inviscid fluid.   This 
reduction (see \cite{STS00}) is done directly at the primary bifurcation to SW and does not rely 
on the symmetry arguments we develop here.

I.  {\bf Simple hexagons}.  These common patterns are favored by quadratic interactions requiring  
harmonic eigenmodes.  We therefore take the coprime integer associated with the instability ($n$, say) 
to be even, and set $N=3$, $\Omega_1=\Omega_2=\Omega_3 = n \omega/2$ in the TW 
expansion (\ref{eq:h}).  The neutral stability curves in the $(k,f)$ plane ($f^2 \equiv |f_m|^2+|f_n|^2$) 
\begin{figure}[ht]
\centerline{\includegraphics[width=2.5in]{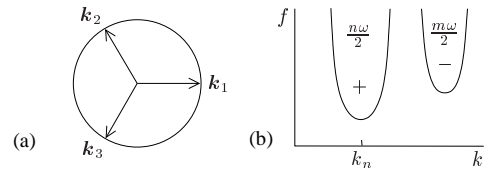}}
\caption{Simplified sketch of when hexagons arise.  With n even the modes excited by $f_n$ are 
harmonic (labeled ``$+$'') and those excited by $f_m$ are subharmonic (``$-$'').}
\label{fig:hex}
\end{figure}
might  resemble the sketch in Fig.~\ref{fig:hex}b.    
Spatial symmetries (translation, reflection, and $60^\circ$ rotation) in combination 
with Eqs.~(\ref{eq:tempsyms}) lead to 
\begin{align}
\dot{Z}_1^+ = &\:\, \upsilon Z_1^+  - \tilde{\mu}f_n Z_1^- 
+ \varpi (Z_2^+\bar{Z}_2^- +Z_3^+ \bar{Z}_3^-)Z_1^- \nonumber \\
& + \left\{\begin{array}{ll} \!\!\tilde{\beta}f_m^{\frac{n}{2}}(\bar{Z}_2^+ 
\bar{Z}_3^- +\bar{Z}_2^- \bar{Z}_3^+) + \tilde{\gamma} \bar{f}_m^{\frac{n}{2}}
\bar{Z}_2^-\bar{Z}_3^-, & m=1\\
\!\!\tilde{\alpha} f_m^{\frac{n}{2}}\bar{f}_n^{\frac{m-3}{2}}\bar{Z}_2^+\bar{Z}_3^+, 
 & m> 1 \end{array}\right\} \nonumber\\
& + Z_1^+ \sum_{j,\,\pm} \tilde{a}_j^\pm |Z_j^\pm|^2. 
\label{eq:TWhex}
\end{align}
The other five equations follow from symmetry.  At each order in amplitude, up to third order, 
the leading order terms (in $f_m$, $f_n$, $\gamma$) have been kept.   Terms of order $|Z|^3 f$ 
are neglected here (and in the examples that follow) because they produce no new qualitative effects.  
Note that the quadratic terms in Eqs.~(\ref{eq:TWhex}) split into two cases: $m=1$ and $m > 1$.   
The time reversal symmetry $\kappa$ in (\ref{eq:tempsyms}) 
forces most of the coefficients to be imaginary at leading order.  For example, we may write
\begin{align}
\tilde{\mu}=\tilde{\mu}_r \gamma + i(\tilde{\mu}_i+\tilde{\mu}_m |f_m|^2+
\tilde{\mu}_n |f_n|^2+\tilde{\mu}_\gamma \gamma^2) + ...\,. \label{eq:tildemu}
\end{align}
Here $\tilde{\mu}_r, \tilde{\mu}_i$, etc.~are real and, with an appropriate sign convention 
for the forcing term in the governing equations, we can take $\tilde{\mu}_i > 0$.  The only 
coefficient which does not submit to a simple transcription of Eq.~(\ref{eq:tildemu}) is $\upsilon$ 
because, having chosen the wavevectors of the expansion to lie on the critical circle 
(itself a function of $\gamma$), we must have the detuning, ${\rm Im}(\upsilon)$, vanish 
at $\gamma = f = 0$.  

\begin{figure}[h]
\centerline{\includegraphics[width= 3.3in]{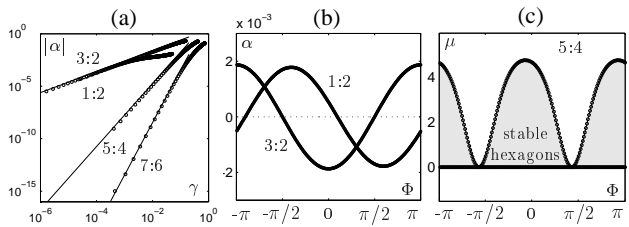}}
\caption{Numerical calculations based on the Zhang Vi\~nals equations (for a summary of the 
procedure see \protect\cite{STS00}). (a) $|\alpha(\gamma)|$ for several choices of $m$:$n$ 
\protect\cite{foot1}.  Solid lines show predicted scalings. (b) $\alpha(\Phi)$ for $m$:$n = 1$:2 and 3:2, 
both with $\gamma \simeq 0.001$.  These cases have identical scaling in $\gamma$ but are phase-shifted in 
$\Phi$.   (c) Range of stable hexagons in Eqs.~(\ref{eq:SWhex}) for $m$:$n = 5$:4 and (realistic) 
parameters such that $\gamma \simeq 0.245$.}
\label{fig:hexnum}
\end{figure}

At the bifurcation to SW, eigenvectors take the form $Z_j^{\pm}=e^{\pm i \varphi}A_j$ 
where $\varphi = \phi_n/2 - \pi/4 + {\cal O}(\gamma)$.  A center manifold reduction yields
\begin{align}
\dot{A}_1 = \mu A_1 \!+ \alpha \bar{A}_2\bar{A}_3 \!+\! A_1\! \left(a|A_1|^2\!+ 
b|A_2|^2\!+ b |A_3|^2\right), \label{eq:SWhex}
\end{align}
with equations for $A_2$ and $A_3$ obtained by cyclic permutation.  The normal form 
coefficients are given (to lowest order) by
\begin{align}
&\alpha = -|f_m|^{\frac{n}{2}}\! \left\{\begin{array}{ll} (2\tilde{\beta}_i-
\tilde{\gamma}_i) \,\sin \Phi, & m=1\! \\
\tilde{\alpha}_i |f_n|^\frac{m-3}{2}\cos \Phi, & m > 1\! \end{array}\right\}\!,
\;\;\; a = C_1\, \gamma, \nonumber \\ 
&b = C_2\, \gamma + \frac{|f_m|^n}{\gamma}\!\left\{\begin{array}{ll}
\:\:C_3\cos^2\Phi, &\!m=1\! \\ \!\!-|C_4| |f_n|^{m-3}\sin^2\Phi, & \!m > 1\! \end{array}\right\}\!.
\end{align}
Here $\Phi = \pi/4-\phi_f/2$ with $\phi_f \equiv m\phi_n-n\phi_m$ (this is the only linear combination 
of $\phi_m$ and $\phi_n$ which cannot be set arbitrarily just by redefining the origin in time).   The 
$C_j$ are known functions of $\upsilon_r$, $\tilde{\mu}_m$, etc., independent of $f_m$, $f_n$, 
$\gamma$ (see \cite{PS02}).   Note that $\alpha$ inherits the $f_m$, $f_n$ dependence of the 
corresponding resonant term in Eqs.~\ref{eq:TWhex} and, since at the bifurcation to SW we have 
$|f_n|, |f_m| \sim \gamma$, it follows that $\alpha$ scales as $\gamma^{n/2}$ if $m=1,3$ and 
$\gamma^{(n+m-3)/2}$ otherwise.   Furthermore, $\alpha$ has a simple harmonic dependence on 
the $T_\tau$-invariant phase $\Phi$.   Both of these predictions are borne out in Fig.~\ref{fig:hexnum}
for several values of $m$:$n$.  This figure also illustrates how the oscillations in $\alpha$ can lead to 
oscillations in the range of forcing amplitude over which hexagons are stable.  Similar oscillations have 
been seen experimentally \cite{EF94,KPG98}.   One can understand such behavior as follows: while the 
linear eigenmodes are phase-locked to the forcing $f_n$ which parametrically excites them, the nonlinear 
forcing from the resonant terms (which involves $f_m$)  attempts to induce a different temporal phase.  
These two influences may cooperate or compete, depending on the relative phase $\phi_f$.

\begin{figure}[ht]
\centerline{\includegraphics[width=2.7in]{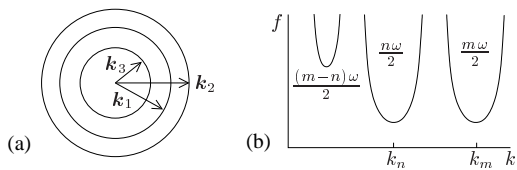}}
\caption{Sketch of the linear problem for 2MS.  We take $m > n$.}
\label{fig:2MS}
\end{figure}

\begin{figure}[h]
\centerline{\includegraphics[width=3in]{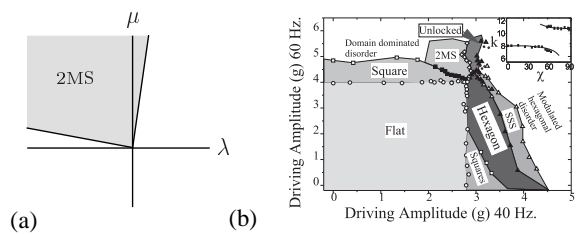}}
\caption{(a) Region of stable 2MS under the assumptions of small damping, Hamiltonian structure, 
and supercriticality, (b) Experimental data, courtesy of J. Fineberg, reproduced from \protect\cite{AF98}.  
Since $\lambda \sim |f_n|-|f_n|^{\rm crit}$ and $\mu \sim |f_m|-|f_m|^{\rm crit}$ the orientation of 
the axes in (a) and (b) is the same.}
\label{fig:2MSbifsets}
\end{figure}

II. {\bf Two-mode superlattices (2MS)}.  These were observed with almost all forcing ratios 
by Arbell and Fineberg \cite{AF98,AF01} near the {\it bi-critical} point where 
the two modes driven by $f_m$ and $f_n$ onset simultaneously.  The relevant three-wave 
interaction also includes a third linearly damped mode (see Fig.~\ref{fig:2MS}b)
with predominant oscillations at half the ``difference frequency''.  The  2MS states typically emerge 
from an underlying square or hexagonal lattice and involve multiple resonant triads \cite{AF98,AF01}.  
For simplicity, we focus here on a single characteristic triad (Fig.~\ref{fig:2MS}a), i.e., we treat an 
invariant subspace of the full 2MS problem.   A combination of spatial and temporal 
symmetries~(\ref{eq:tempsyms}) again determines the form of the six TW evolution equations.   
In the vicinity of the bi-critical point these TW equations reduce to SW equations
\begin{align}
&\dot{A}_1 = \lambda A_1 + A_1\left(a|A_1|^2+b|A_2|^2\right),\nonumber \\
&\dot{A}_2 = \mu A_2 + A_2\left(c|A_1|^2+d|A_2|^2\right),
\label{eq:Aeq2MS}
\end{align}
with coefficients of the form
\begin{align}
&\left(\begin{array}{c} a \\ b \\ c \\ d \end{array}\right) = \left(\begin{array}{c} C_1 \\ C_2 \\ C_3 
\\ C_4 \end{array}\right)\gamma + \left(\begin{array}{c} 0 \\ |C_5|  \tilde{\beta}_i \tilde{\kappa}_i \\ 
\!\!\!\! -|C_5| \tilde{\alpha}_i \tilde{\kappa}_i  \\ 0 \end{array}\right) \frac{1}{\gamma}\,.
\label{eq:2MScoef}
\end{align}
Here the $C_j$ are known functions, independent of $\gamma$, $f_m$, and $f_n$ while 
$\tilde{\alpha}_i$, $\tilde{\beta}_i$, and $\tilde{\kappa}_i$ are the imaginary parts of the quadratic 
resonant coefficients in $\dot{Z}_1^+$,  $\dot{Z}_2^+$, and $\dot{Z}_3^+$, respectively.   The 
most important effect of Hamiltonian structure on the TW equations comes via these last three 
coefficients.   Specifically, if at $\gamma=0$ there exists a Hamiltonian ${\cal H}$ such that 
$d Z_j^\pm /d t = \mp i \partial {\cal H}/\partial \bar{Z}_j^\pm$, then $\tilde{\alpha}_i = 
\tilde{\beta}_i = \tilde{\kappa}_i$.  The equality here is an overstatement since the equations 
(and hence the coefficients) can always be rescaled.   One expects simply $\tilde{\alpha}_i \sim 
\tilde{\beta}_i \sim \tilde{\kappa}_i$ and thus, for small damping, $1 \ll b \sim -c$.   If, in addition, 
both primary bifurcations (to pure modes) within Eqs.~(\ref{eq:Aeq2MS}) are supercritical then 
$0 < -a \sim -d \ll 1$.  Under these conditions stable 2MS states (mixed states within 
Eqs.~(\ref{eq:Aeq2MS}))  inhabit the region shown in Fig.~\ref{fig:2MSbifsets}a.  This region 
agrees quite well with the experiments of Arbell and Fineberg \cite{AF98} despite 
being based on a simplified version of the 2MS states.  The indifference of Eqs.~(\ref{eq:2MScoef}) 
to $m$ and $n$ helps explain why 2MS patterns were observed for virtually all forcing 
ratios \cite{AF98}.

\begin{figure}[h]
\centerline{\includegraphics[width = 2.5 in]{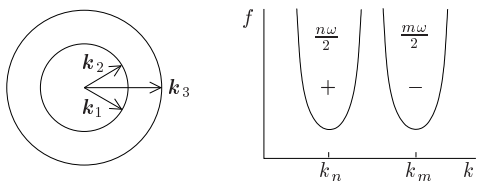}}
\caption{Sketch of the linear picture for 2MR.  We take m odd and n even.}
\label{fig:2kR}
\end{figure}

III.  {\bf Two-mode rhomboids (2MR)}.  These patterns are found \cite{AF00} near the bi-critical point 
but, unlike 2MS, do not rely on damped modes; see Fig.~\ref{fig:2kR}.
As before, spatial symmetries and Eqs.~(\ref{eq:tempsyms}) determine the form of the 
six TW equations that  reduce to SW equations at the bi-critical point:
\begin{align}
\dot{A}_1 &\!= \lambda A_1 \!+ \alpha \bar{A}_2 A_3 \!+\! 
A_1 \big(a|A_1|^2 \!+ b|A_2|^2 \!+ c|A_3|^2\big), \nonumber \\
\dot{A}_3 &\!= \mu A_3 \!+ \beta A_1 A_2 \!+\! A_3\big(d|A_1|^2 \!+ d |A_2|^2 \!+ e|A_3|^2\big), 
\label{eq:Aeq2kR}
\end{align}
with  resonant coefficients of the form
\begin{align}
\left(\begin{array}{c} \alpha \\ \beta \end{array}\right) = \left(\begin{array}{c} \,\tilde{\alpha}_i \\ 
\!\!\!-\tilde{\beta}_i \end{array}\right)|f_m|^{\frac{n-2}{2}}|f_n|^{\frac{m-1}{2}}\cos \Phi.
\end{align}
Again $\Phi = \pi/4-\phi_f/2$ with $\phi_f = m\phi_n-n\phi_m$.  Hamiltonian structure at 
$\gamma=0$ implies $\tilde{\alpha}_i \sim 
\tilde{\beta}_i$ and thus $\tilde{\alpha}_i \tilde{\beta}_i > 0$, or equivalently, $\alpha \beta < 0$ 
(see \cite{C01} for proof of an analogous relation in a class of self-adjoint hydrodynamic problems).  
The condition $\alpha \beta < 0$ has enormous dynamical consequences for Eqs.~({\ref{eq:Aeq2kR}}) 
and is associated  with the presence of modulated drifting waves and heteroclinic cycles 
\cite{tripleref,PS02}.   We have confirmed the opposing signs of $\alpha$ and $\beta$, as well as the 
predicted scaling $|\alpha|$, $|\beta| \sim \gamma^{(m+n-3)/2}$, for a wide range of forcing ratios 
by numerically calculating these coefficients from the Zhang-Vi\~nals equations.  The exponential 
decrease of $|\alpha|$ and $|\beta|$ with $m+n$ helps explain why 2MR have been seen in 
\cite{AF00,AF01} for $m$:$n$ = 3:2 and 5:4 but not with higher values of $m+n$; for $m$:$n$ = 1:2 
the same three-wave resonance is observed, but as part of a superlattice-II type state \cite{AF01}.

In conclusion, we have shown that with weak damping the ``broken'' symmetries of time translation, 
time reversal, and Hamiltonian structure impose significant constraints on the equations describing 
SW near onset.   These include scaling laws of resonant interactions with $f_m$ and $f_n$, 
harmonic dependence on $\phi_f$, and the dynamically significant preference for resonant 
coefficients of opposite sign in the 2MR problem.   We have verified numerically these 
predictions for the Zhang-Vi\~nals equations, and used them to interpret experimental results - they 
explain the oscillations seen with varying temporal phase in \cite{EF94,KPG98}  and predict the correct 
region of parameter space for 2MS states.   We note, however, that it is not always 
clear whether the damping  in a given experiment is ``small enough'' for our results to apply.  This is 
particularly true when shallow containers are used and the damping due to boundary layer effects is 
important.  It is unclear as well whether some procedures (such as the simple Taylor expansion in 
$\gamma$) would carry over to the full problem (Navier-Stokes equations with realistic boundary 
conditions) where the addition of damping constitutes a singular perturbation.  

Finally, we emphasize that the ideas developed in this letter can be fruitfully applied to a variety 
of other patterns, and could be extended to include more general periodic forcing functions.  
In particular, they indicate how one might control resonant wave interactions (and hence pattern 
formation) through a judicious choice of the harmonic content of the forcing function.   Furthermore, 
many aspects of our mathematical framework, developed here for parametrically excited SW 
patterns, should carry over to other problems in which TW naturally occur, e.g.~via Hopf 
bifurcation.  It would be interesting, for instance, to apply our framework to the problem 
of SW pattern control via weak temporal modulation of the control parameter 
\cite{RCK88,D00}.

\begin{acknowledgments}
We thank J. Fineberg, M. Golubitsky, H. Riecke, C. Topaz, and P. Umbanhowar for helpful discussions.  
This work was supported by NASA grant NAG3-2364 and NSF grant DMS-9972059.
\end{acknowledgments}

\bibliography{PS02}

\begin{thebibliography}{16}
\bibitem[{foo()}]{foot1}
\bibinfo{note}{For all $m$:$n$ pairs tested (more than 10) the numerical
  calculations confirm our predictions with the exception of 3:4, where the
  term given in Eqs.~(7) seems to be absent, giving way to one of higher order.
  This disagreement could be due to inadequate numerical resolution, or to a
  hidden symmetry of some kind.}
\end{thebibliography}

\end{document}